# Lane–Emden Equation: Picard vs Pade

**Zakir F. Seidov, Research Institute, College of Judea and Samaria,**

**Ariel , 44837 Israel**

The approximate analytical solutions are found to Lane-Emden equation using Picard type iteration and rational Pade approximants. The presented values for the case of *n=2* are more accurate than any one known in literature.

## Introduction

The investigation of Lane – Emden (LEE) equation has a long – standing history (see, e.g., [1 – 10], to mention only part of a vast literature). Recently Schaudt [6] has shown that Picard type iteration scheme may be used to show the existence, uniqueness and regularity of global solutions of LEE, in particular, for $n \geq 1$.

In this paper it was shown that straightforward usage of Picard iteration at general case of arbitrary $n$ (even for $n \geq 1$) is not possible analytically, and so we apply this procedure for particular case of $n = 2$. The iterations are converging and allow to obtain the solution up to first zero of solution with great accuracy.

Then we consider the modification of iteration scheme allowing to get very accurate Pade approximations and the values of relevant parameter at first zero of Lane – Emden function.

We conclude that integral form of LEE allows to get (analytical) approximations in the terms of economized series (rather polynomials) and of Pade approximants in the way which is not straightforward from differential form of LEE.

Presented results, for the case of $n = 2$, are more accurate than ones known in literature.

### ODE and integral equation

Lane – Emden equation (LEEn) is a second order ODE with parameter n [1, 2]:

$$y''(x) + \frac{2}{x} y'(x) = -y^n(x), \quad y(0) = 1, \quad y'(0) = 0. \quad (1)$$

By usual way (see, eg., [3 – 6]) it may be rewritten as nonlinear integral equation of Volterra type:

$$y(x) = 1 - \int_0^x s \left(1 - \frac{s}{x}\right) y^n(s) \, ds. \quad (2)$$

This integral form of LEEn is more profitable comparing to (1) in several aspects, reason for this being the explicit taking into account the initial conditions while deriving (2) from (1).

### Iterative procedure in general case

We use here (2) for obtaining approximate analytical solutions of LEEn.
To this end, we rewrite (2) as:

$$y_{k+1}(x) = 1 - \int_0^x s \left(1 - \frac{s}{x}\right) y_k{}^n(s) \, ds. \quad (3)$$

In this form, one clearly sees the possibility to solve the LEEn by Picard type iteration procedure, which should be converging at least for $n \geq 1$ [6].

Note that there is no such iterative procedure known for LEE in form (1).
Now we start with $y_0(s) = 1$. We get (for any $n$!):

$$y_1(x) = 1 - \frac{x^2}{6}, \quad (4)$$

$$y_2(x) = \frac{-2+n}{1+n} + \frac{3\left(1-\frac{x^2}{6}\right)^{n+1}}{(1+n)} + \frac{1}{3}x^2 \,_2F_1\left(\frac{3}{2}, -n; \frac{5}{2}; \frac{x^2}{6}\right). \quad (5)$$

Next step leads to uncalculable integrals thus precluding any further (analytical!) iteration for general $n$, starting with ideal zero-step $y_o(x) = 1$ (in the spirit of [6]).

### Case of n = 2

So we consider particular values of $n$, e.g. $n = 2$:

$$y_2(x) = 1 - \frac{x^2}{6} + \frac{x^4}{60} - \frac{x^6}{1512}. \quad (6)$$

Note that, in (6), only first three terms coincide with power series expansion of $y(x)$ around $x = 0$.
Rather each Picard iteration gives in some sense economized power series of the order which rapidly increases with iteration number.
  This allows to calculate function $y(x)$, and even the first zero of $y(x)$, with great accuracy.
  Second iteration, $y_2(x)$ (6), has the first zero $X_2 = 3.89295$, which is a rather rough estimation. In the Table 1, we present the values of first zero, $X_k$, such that $y_k(X_k) = 0$, (for $n = 2$!) from several Picard iterations:

**Table 1. Parameters at first zero of solution to LEE2**

| k | $X_k$ Picard | $X_{p,2k-2}$ Pade | $\mu = -x^2 y'(x) /. x \to X_{p,2k-2}$ Pade |
|---|---|---|---|
| 0 | – | – | – |
| 1 | $6^{1/2}$ | – | – |
| 2 | 3.89295362 | 3.87298173844 | 3.0983859052 |
| 3 | 4.48977893 | 4.40920514416 | 2.2244983076 |
| 4 | 4.33734237 | 4.35230464399 | 2.4143159741 |
| 5 | 4.35410947 | 4.35287035336 | 2.4110793941 |
| 6 | 4.35280006 | 4.35287444601 | 2.4110474718 |
| 7 | 4.35287815 | 4.35287459842 | 2.4110459822 |
| 8 | – | 4.35287463348 | 2.4110465354 |
| 9 | – | 4.35287459594 | 2.4110460122 |
| 10 | – | 4.35287459595 | 2.4110460121 |
| 11 | – | 4.35287459595 | 2.4110460121 |

### Picard iteration, power series and Pade approximation

We observe the convergence of Picard type iteration scheme of solution to LEEn starting with zero approximation $y_0(x) = 1$, for LEEn of parameter $n = 2$.
The reason of this convergence, one may guess, is that in Picard iterations, even in this rather simple case of LEE2, we get not exact power series, but in some sense economized power series.
  It is known [7], that exact power series expansion of function $y(x)$ for arbitrary $n$ has radius of convergence in all cases less than first zero of $y(x)$. So method of power series can not allow to find first zero of function $y(x)$. However one can use power series as an intermediate result to get from them Pade approximations, which are (much more) accurate than corresponding parent power series (see, e.g., [8, 9]).
  In order to illustrate the advantage of Pade approximation in comparison with Picard iterative scheme
 (not speaking of pure power series with their small radius of convergence!)
we present so called diagonal Pade approximation of eighth order, PA[8, 8], to solution of LEE2:

$$PA[8, 8] = (838857127878000 - 36744139139400\, x^2 - 296471637000\, x^4 - 4578235200\, x^6 - 39340559\, x^8) / (6\, (139809521313000 + 17177563695600\, x^2 + 483356654550\, x^4 - 3069772860\, x^6 - 18829469\, x^8)). \quad (7)$$

It's size is incomparably smaller than that of Picard iterations of the similar accuracy of evaluated value of first zero $X_{p,8} = 4.35287035336$.

### Pade approximation

So we suggest the more effective modification of Picard iteration scheme.
First we find power series using *expansion* of $y^n(x)$ in the integrand in (3) and then calculate the integral. We note that the general recursive relation for coefficients of power series for function $y(x)$ of arbitrary $n$ is given in [7]. Here we wish to underline that even not knowing these recurrence relations one can calculate the power series coefficients with iterative scheme from (3) while obtaining the coefficients of power series from (1) is not so straightforward.

We start with $y_1(x) = 1 - \frac{x^2}{6}$.

Then using iterative procedure (3) several times we get the power series for general $n$:

$$y_s(x) = 1 - \frac{x^2}{6} + \frac{n\, x^4}{120} - \frac{n\,(-5 + 8n)\, x^6}{15120} + \frac{n\,(70 - 183n + 122n^2)\, x^8}{3265920} - \frac{n\,(-3150 + 10805n - 12642n^2 + 5032n^3)\, x^{10}}{1796256000} + \frac{n\,(138600 - 574850n + 915935n^2 - 663166n^3 + 183616n^4)\, x^{12}}{840647808000} - \frac{n\, x^{14}}{1235752277760000}(-21021000 + 101038350n - 199037015n^2 + 200573786n^3 - 103178392n^4 + 21625216n^5) + \frac{n\, x^{16}}{1008373858652160000}(1891890000 - 10267435500n + 23780949500n^2 - 30057075285n^3 + 21827357636n^4 - 8618115372n^5 + 1442431856n^6). \quad (8)$$

Using this power series we get several Pade[k, k] approximations:

$$PA[2, 2] = \frac{60 + (-10 + 3n)\, x^2}{60 + 3n\, x^2}; \quad (9)$$

$$PA[4, 4] = (45360\,(-50 + 17n) + 420\,(1250 - 951n + 178n^2)\, x^2 + (-24500 + 29100n - 10849n^2 + 1290n^3)\, x^4) / (45360\,(-50 + 17n) + 420\,(350 - 645n + 178n^2)\, x^2 + 15\,(190n - 321n^2 + 86n^3)\, x^4); \quad (10)$$

$$PA[6, 6] = \frac{(a_0 + a_2\, x^2 + a_4\, x^4 + a_6\, x^6)}{7\,(b_0 + b_2\, x^2 + b_4\, x^4 + b_6\, x^6)},$$

$a_0 = 389188800\,(12372500 - 28353900n + 21372605n^2 - 6679392n^3 + 749132n^4)$,

$a_2 = 2494800\,(-592900000 + 1624109100n - 1644080680n^2 + 792866757n^3 - 184792168n^4 + 16782156n^5)$,

$a_4 = 22680\,(5780040000 - 18348690500n + 23045300300n^2 - 14873877965n^3 + 5246107880n^4 - 963525716n^5 + 72229136n^6)$,

$a_6 = (-3053351700000 + 10729693415000n - 15644831683500n^2 + 12377539036650n^3 - 5750486943975n^4 + 1567735653534n^5 - 231903413748n^6 + 14339949064n^7)$,

$b_0 = 55598400\,(12372500 - 28353900n + 21372605n^2 - 6679392n^3 + 749132n^4)$,

$b_2 = 3207600\,(-30135000 + 98545300n - 120932550n^2 + 68800285n^3 - 18368304n^4 + 1864684n^5)$,

$b_4 = 3240\,(807765000 - 3857983500n + 7146037250n^2 -$

$$6578113455\, n^3 + 3170490776\, n^4 - 762978732\, n^5 + 72229136\, n^6),$$
$$b_6 = n\,(27533555000 - 124961404500\, n + 221524322250\, n^2 - 196938366435\, n^3 +$$
$$92676531282\, n^4 - 21931005924\, n^5 + 2048564152\, n^6).\quad(11)$$

Note that first two Pade approximations, PA[2, 2] and PA[4, 4],
see (9) and (10), are given first in [8],
where they are denoted as $\Theta_n^{[1,1]}(\xi)$ and $\Theta_n^{[2,2]}(\xi)$ respectively,
see Eq. (5) in [8]. The third Pade approximation PA[6, 6], formula (11),
(and forth approximation given in *Appendix*) for LEEn solution of any
parameter *n* are presented here at first. Also it can be shown that all
four Pade approximations reduce to $1 - \frac{1}{6}\,x^2$ at $n \to 0$. Also at cases
$n = 1$ and $n = 5$ the formulas (9-11) and in *Appendix*
provide Pade approximations of known functions:

$$\frac{sin(x)}{x}\ (n=1),\ \text{and}\ \left(1+\frac{x^2}{3}\right)^{-1/2}\ (n=5).$$

In the last column of Table 1 the values of another parameter useful in the theory
of polytropic stars (see [1, 2]) are also given according to Pade approximations.
We note that values in Table 1 are of the largest accuracy known
in literature (in [8] the values of X and $\mu$ are given with 7 decimals).
The one important aspect of these calculations is using an ability
of *Mathematica* [11] of *exact* calculations with *integers*,
and obtained results in the last lines of Table 1 are even more
exact and we omitted several last decimals in presented results.

### Appendix

$$PA[8,8] = \frac{(c_0 + c_2\,x^2 + c_4\,x^4 + c_6\,x^6 + c_8\,x^8)}{7\,(d_0 + d_2\,x^2 + d_4\,x^4 + d_6\,x^6 + d_8\,x^8)};$$

$c_0 = 2778808032000\,(-40539499623000000 +$
$219612227381050000\, n - 511477654026990000\, n^2 + 676587517878560000\, n^3 -$
$562530934795015800\, n^4 + 305353039511462775\, n^5 - 108282312996133392\, n^6 +$
$24204097219027016\, n^7 - 3097602512909376\, n^8 + 173153049930352\, n^9);$

$c_2 = 2724321600\,(15975797005665000000 - 93812750271259250000\, n +$
$241429498430852550000\, n^2 - 360556559054312350000\, n^3 + 347275338859032654000\, n^4 -$
$225793301183157240975\, n^5 + 100428697440662407020\, n^6 - 30180690880628440984\, n^7 +$
$5866332008193485952\, n^8 - 666264385302502832\, n^9 + 33596198313370944\, n^{10});$

$c_4 = 162162000\,(-3482471381558400000 + 21992717390680260000\, n -$
$61813942618784237000\, n^2 + 102517374377946558800\, n^3 -$
$111796633942257211900\, n^4 + 84297992714278241818\, n^5 -$
$44872278588599817018\, n^6 + 16863657707312025135\, n^7 - 4384808086214750418\, n^8 +$
$751272029582714617\, n^9 - 76348442767513352\, n^{10} + 3487415490432569\, n^{11});$

$c_6 = 83160\,(33354537418818000000 - 224014013798829450000\, n +$

$$678331754580949297500000\, n^2 - 1228890803236775724500000\, n^3 +$$
$$1486880953522329325400000\, n^4 - 1267277965093877939470000\, n^5 +$$
$$780433629609239374182250\, n^6 - 349849023783140017725000\, n^7 +$$
$$113257580848517762336160\, n^8 - 25812954742546772620960\, n^9 +$$
$$3930277929213659041120\, n^{10} - 358880445956158112000\, n^{11} + 14860829102968839680\, n^{12});$$

$c_8 = -3935358993175533000000000 + 27489086733745765800000000\, n -$
$8758706664902253834250000\, n^2 + 16905206976827496380025000\, n^3 -$
$22094294289195176033545000\, n^4 + 20667373216468925656750500\, n^5 -$
$14237742823916382438236725\, n^6 + 7311126540259539569417512 5\, n^7 -$
$2796451389786890303764775 0\, n^8 + 7861679390615108424720096\, n^9 -$
$1577321939775971114089424\, n^{10} + 213657036563976016380 96\, n^{11} -$
$174875227939075520143 04\, n^{12} + 652628896248581615 36\, n^{13};$

$d_0 = 2778808032000\, (-40539499623000000 + 219612227381050000\, n -$
$511477654026990000\, n^2 + 676587517878560000\, n^3 -$
$562530934795015800\, n^4 + 305353039511462775\, n^5 - 108282312996133392\, n^6 +$
$24204097219027016\, n^7 - 3097602512909376\, n^8 + 173153049930352\, n^9);$

$d_2 = 8172964800\, (3028027356585000000 - 18826223872160250000\, n +$
$51492765748754750000\, n^2 - 81845560338319050000\, n^3 + 83881693314626656000\, n^4 -$
$57961094822069523075\, n^5 + 27340234743773243460\, n^6 - 8688664784464616088\, n^7 +$
$1779913193666297344\, n^8 - 212276122271447664\, n^9 + 11198732771123648\, n^{10});$

$d_4 = 32432400\, (-4694642010135000 0000 + 3378796996341045 00000\, n -$
$10848040998415820500 00\, n^2 + 20535502296631987 00000\, n^3 -$
$2549884065663832883 000\, n^4 + 21821807406306334 02950\, n^5 -$
$1313346140402699046975\, n^6 + 555572535897326622972\, n^7 - 161765775591138321896\, n^8 +$
$30859692537952223456\, n^9 - 3470706639638745712\, n^{10} + 174370774521628480\, n^{11});$

$d_6 = 83160\, (28393643529405000 0000 - 247121578787511 7500000\, n +$
$959613285379609875 0000\, n^2 - 2204631828670459 3700000\, n^3 +$
$33457928005411825895000\, n^4 - 35407121047497988629250\, n^5 +$
$26826302480245895755525\, n^6 - 14669178502968121829460\, n^7 +$
$5745112561862418989088\, n^8 - 1571007418300434014512\, n^9 +$
$284607051855194912368\, n^{10} - 30664360468491604352\, n^{11} + 1486082910296883968\, n^{12});$

$d_8 = 9\, n\, (17434185082476750000000 - 14664762574761479250000 0\, n +$
$55181489087135183325000 0\, n^2 - 1231952674494568458250000\, n^3 +$
$1822280070039019266490000\, n^4 -$
$1885364478216564017185250\, n^5 + 1400770499278411609550225\, n^6 -$
$753286798160886430110090\, n^7 + 290902568097394058143104\, n^8 -$
$78620673835912275662576\, n^9 + 14104705308862137378704\, n^{10} -$
$1507180671042570654496\, n^{11} + 72514309958317573504\, n^{12}).$